\documentclass[fleqn,usenatbib]{mnras}

\usepackage{newtxtext,newtxmath}
\usepackage[T1]{fontenc}

\DeclareRobustCommand{\VAN}[3]{#2}
\let\VANthebibliography\thebibliography
\def\thebibliography{\DeclareRobustCommand{\VAN}[3]{##3}\VANthebibliography}

\usepackage{graphicx}	
\usepackage{amsmath}	
\usepackage{subfigure}
\title{Investigating the multi-drifting behavior of subpulses in PSR J2007$+$0910 with the FAST}

\author[X. Xu \& L. H. Shang et al.]{
Xin Xu,$^{1,3}$
Lunhua Shang,$^{2,3}$\thanks{E-mail: lhshang@gznu.edu.cn (SLH)}
Qijun Zhi,$^{2,3}$\thanks{E-mail: qjzhi@gznu.edu.cn (ZQJ)}
Shijun Dang,$^{2,3}$\thanks{E-mail: dangsj@gznu.edu.cn (DSJ)}
Shi Dai,$^{4}$
Jie Tian,$^{2,3}$
Yan Yu,$^{2,3}$
\newauthor
Qingying Li,$^{2,3}$
Shuo Xiao,$^{2,3}$
Dandan Zhang,$^{2,3}$
\\
$^{1}$School of Mathematical Sciences, Guizhou Normal University, Guiyang 550001, China.\\
$^{2}$School of Physics and Electronic Science, Guizhou Normal University, Guiyang 550001, China.\\
$^{3}$Guizhou Provincial Key Laboratory of Radio Astronomy and Data Processing, Guizhou Normal University, Guiyang 550001, China.\\
$^{4}$School of Science, Western Sydney University, Locked Bag 1797, Penrith South DC, NSW 2751, Australia.\\
}

\date{Accepted XXX. Received YYY; in original form ZZZ}

\pubyear{2023}

\begin{document}
\label{firstpage}
\pagerange{\pageref{firstpage}--\pageref{lastpage}}
\maketitle

\begin{abstract}
The multi-drifting subpulse behaviors in PSR J2007$+$0910 have been studied carefully with the high sensitivity observations of the Five-hundred-meter Aperture Spherical radio Telescope (FAST) at 1250~MHz. 
We found that there are at least six different single emission modes in  PSR J2007$+$0910 are observed, four of which show significant subpulse drifting behaviors (modes A, B, C, and D), and the remaining two (modes $E_1$ and $E_2$) show stationary subpulse structures. 
The subpulse drifting periods of modes A, B, C, and D are $P_{3, A} = 8.7 \pm 1.6~P$, $P_{3, B} = 15.8 \pm 1.2~P$, $P_{3, C} = 21.6 \pm 1.3~P$ and $P_{3, D} = 32.3 \pm 0.9~P$, respectively, where $P$ represents the pulse period of this pulsar. 
The subpulse separation is almost the same for all modes $P_2 = 6.01 \pm 0.18 ^\circ$.
Deep analysis suggests that the appearance and significant changes in the drifting period of multi-drifting subpulse emission modes for a pulsar may originate from the aliasing effect.
The observed non-drifting modes may be caused by the spark point move with a period $\sim P_2$.
Our statistical analysis shows that the drift mode of this pulsar almost always switches from slower to faster drifts in the mode change.
The interesting subpulse emission phenomenon of PSR J2007$+$0910 provides a unique opportunity to understand the switching mechanism of multi-drift mode. 
\end{abstract}



\section{introduction}\label{sec_1}
Pulsar is highly magnetized neutron star and known for its highly periodic pulse emission, and was first discovered in 1968 \citep{1968Natur.217..709H}. 
In the decades since their discovery, more than three thousand pulsars have been discovered \citep{2005AJ....129.1993M, 2021RAA....21..107H}. 
During this half-century of observational research, more and more interesting phenomena, such as subpulse drifting, mode changing, giant pulses, and the periodic behavior of micropulse, have been discovered.
Many theoretical models have been proposed to explain these phenomena \citep[e.g. ][]{1971ApJ...164..529S, 1975ApJ...196...51R, 1998A&A...333..172Q,2004ApJ...616L.127Q}.
However, there is still no consensus on its radiation mechanism. This requires more observation samples for statistical research to limit the theoretical model.

Generally, the integrated pulse profile is the result of hundreds or even thousands of single pulses accumulated and summed at the corresponding phase.
For most pulsars, the averaged pulse profile  is very stable in time, and its morphology is related to the pulse width and frequency of observation \citep{1983ApJ...274..333R, 2021ApJ...917..108X,2017MNRAS.468.4389S,2021ApJ...916...62S, 2022ApJ...926...73Z}. 
In contrast, for most pulsars, the single pulse composed of one or more subpulses, the morphology and intensity in a periodic radiation window often changes randomly. 
However, for some pulsars, the phase of subpulse changes regularly, the so-called drifting subpulse. 
The drifting subpulse phenomenon was firstly observed by \citet{1968Natur.220..231D}.
The drifting subpulse usually shows a regular shift forward or backward in phase.
Up to now, the drifting subpulse phenomenon has been observed in hundreds of pulsars with the advancement of observation technology \citep{2006A&A...445..243W, 2007A&A...469..607W, 2019MNRAS.482.3757B}. This implies that the subpulse drift phenomenon may be a common behavior in pulsars, and the samples of subpulse drifting pulsars will be greatly expanded with the establishment and observation of giant telescopes, which makes it possible to carry out in-deep statistical research on the subpulse drift phenomenon of large samples in the future.

Subpulse drift can be represented by three parameters, which are the vertical band spacing at the phase of the same pulse ($P_3$), the horizontal time interval between successive drift bands ($P_2$), and the drift rate ($\Delta \phi = P_2 / P_3$).
\citet{1975ApJ...196...51R} gives an early theoretical explanation of this phenomenon. 
they suggested that localized pockets of quasi-stable electrical activity called sparks exist near the pulsar’s surface.
These sparks move relative to the polar cap surface about the magnetic axis due to the $E \times B$ drift, it's like a slightly rotating carousel, which is known as the carousel model. 
The radiation beam will also have sub-beam that reflect these sparks. 
The observed drift behavior is produced when the line of sight sweeps through these rotating sub-beam.
The behavior of the drift bands is determined by the carousel rotation rate, $P_4$, and the number of sparks in the carousel, $n$ \citep{2019ApJ...883...28M}.

It is worth noting that some pulses show some interesting characteristics of pulse drift.
Examples include frequent switching of multiple drift modes \citep{1970ApJ...162..727H, 2016A&A...590A.109W, 2019ApJ...883...28M, 2023MNRAS.520.1332Z}, subpulse "memory" across nulls \citep{1983MNRAS.204..519L, 2022MNRAS.509.2507R}, bi-drifting \citep{2004ApJ...616L.127Q, 2020ApJ...896..168S, 2022RAA....22b5018S},  abrupt transitions of drift direction \citep{1985MNRAS.215..281B, 2022ApJ...934...23S}, etc.
All these particular phenomena challenge the traditional carousel model, which means that the model should be extended \citep{2006ChJAS...6b..18E}.
Several studies have shown connections between nulling, subpulse drift and mode changing. 
For example, the speeding up or slowing down of subpulse drift after nulling \citep[e.g. ][]{2004A&A...425..255J, 2017ApJ...850..173G}. 
The change in subpulse drift whether  is caused by nulling needs more in-depth study. 
The study of multi-drift subpluse pulsars exhibiting no null pulse provides an opportunity to understand this problem.

Pulsar J2007$+$0910 was discovered in the Arecibo drift-scan searches \citep{2005MNRAS.363..929C}. 
Its period is $0.459$ s and a $\dot{P}$ of $0.332 \times 10^{-15}s\,s^{-1}$. 
Previously, \citet{2005MNRAS.363..929C} has studied the single pulse of this pulsar by Arecibo at 430~MHz, found its subpulse drift behavior, and calculated the period modulation of the subpulse to be $P_3 = 11.8~P$, where $P$ is the pulse period.
Recently \citet{2023MNRAS.520.4562S} measured the period modulation of this pulsar $P_3 = 21~P$ using MeerKAT at 1284~MHz.
There are differences in their results. There is therefore a great necessity to study this pulsar in more detail using highly sensitive telescopes and to obtain its exact modulation period.

In this paper, we used the high-sensitivity observations at center frequency 1250~MHz with a 400~MHz bandwidth of the Five-hundred-meter Aperture Spherical Radio Telescope (FAST) of China to study the drifting subpulse behavior in PSR J2007$+$0910. 
FAST is currently the biggest single-dish radio telescope in the word \citep{2006ScChG..49..129N}, and because of its extraordinarily high sensitivity, a wide range of fascinating phenomena have been discovered \citep[e.g. ][]{2023MNRAS.520.1332Z, zhangdd2023, 2023ApJ...949..115Y}.
The observation of FAST will reveal a more detailed radiation structure of pulsars, thereby providing strict limitations for theoretical research on single pulse radiation of pulsar. 
The paper is structured as follows: In Section\,\ref{sec_2}, the FAST observations of PSR J2007$+$0910 and data reduction will be given. 
The bshaviors of single pulse and subpulse and the magnetosphere configuration of PSR J2007$+$0910 will presented in Section\,\ref{sec_3} and Section\,\ref{sec_4}, respectively.
The implication of our results are discussion in Section\,\ref{sec_5}
Finally, We concluded with a short summary in Section\,\ref{sec_6}.

\section{observations and reduction}\label{sec_2}
\begin{figure} 
 \centering
\includegraphics[width=0.48\textwidth]{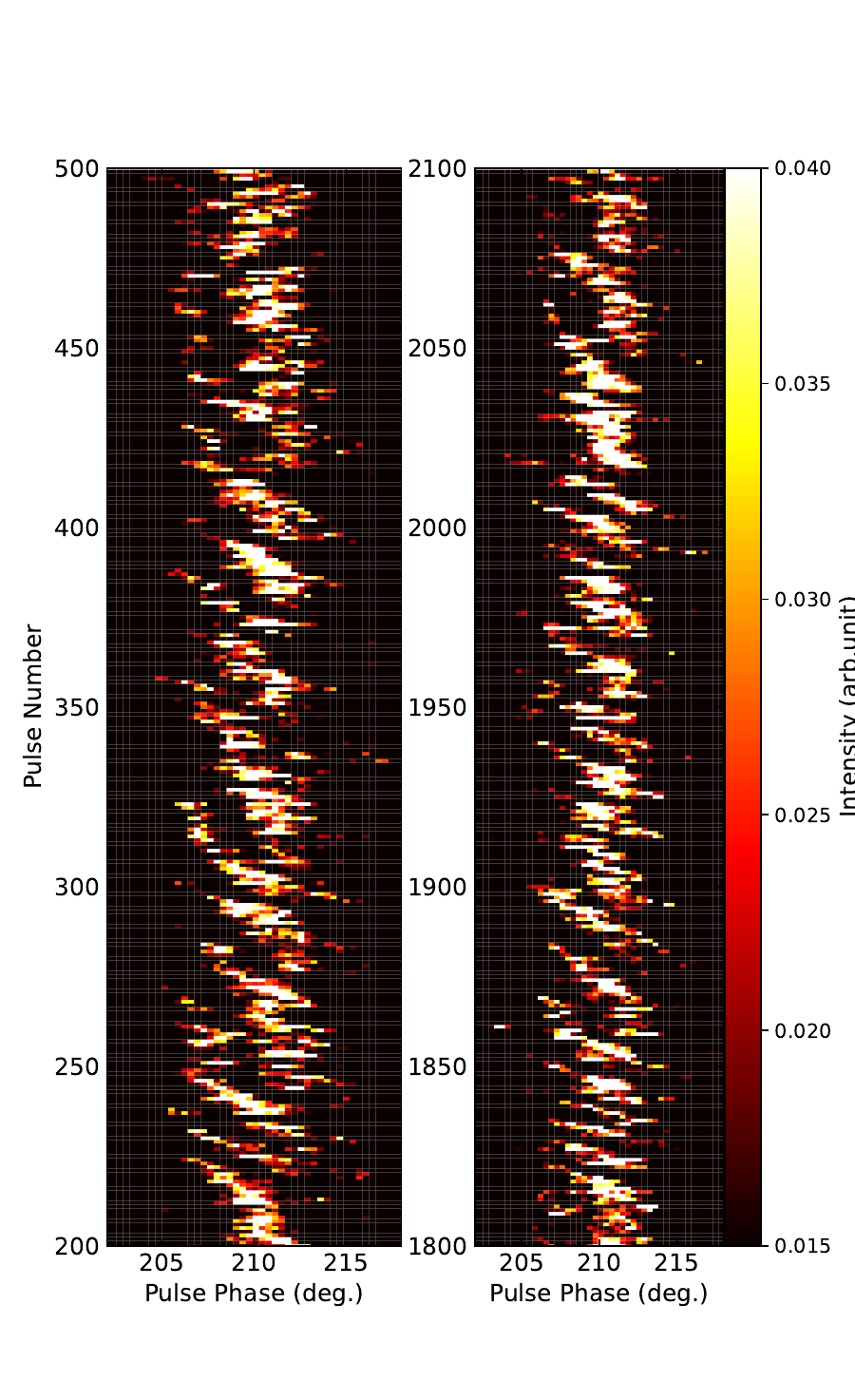}
 \caption{Two single pulse plots of J2007$+$0910 with 300 consecutive pulses in each plot. An example of fast switching between different drift modes is shown. Four drifting modes and two non-drifting mode are included. A specified saturation intensity range is used to highlight the drift phenomena of the single pulses.
 \label{fig_1}} 
\end{figure}

FAST is a Chinese mega-science project. 
It is located in Guizhou province, China. 
The longitude and latitude of the FAST on the earth are 106.9$^{\circ}$E and 25.7$^{\circ}$N, respectively.
The designed total aperture is 500 meters, but the designed effective aperture is 300 meters.
The main structure of the FAST was completed on September 25, 2016, and entered the commissioning phase\,\citep{2019SCPMA..6259502J, 2020RAA....20...64J}.
During the early commissioning phase from September 2016 to May 2018, an ultra-wideband receiver was used at FAST.
After May 2018, a 19-beam receiver with a bandpass of 1.05--1.45\,\rm{GHz} has been used in its observation.
In this paper, the observations were carried out with the central beam of FAST 19-beam receiver on 2020 October 13 (MJD 59135). 
The duration of 19-beam observation is 2400\,\rm{s}.  
The observing frequency range is from 1050~MHz to 1450~MHz with 4096 channels, and the time resolution is 49.152 $\,\rm{\mu}s$. 
Data were recorded in the search mode and in PSRFITS format \citep{2004PASA...21..302H}.
With the timing ephemeris provided by the ANTF pulsar catalog \citep[PSRCAT]{2005AJ....129.1993M}, we used the DSPSR package \citep{2004PASA...21..302H,2011PASA...28....1V} to fold the data, and obtained the single pulse profiles. 
In observation,  radio signals from pulsars are usually more or less contaminated by narrowband non-pulsar radio radiation at a certain frequency.
These narrowband non-pulsar radio-frequency interference (RFI) are flagged and removed by using the software package PSRCHIVE \citep{2004PASA...21..302H, 2012AR&T....9..237V}.
Through separate measurements of a noise diode signal injected at an angle of $45 ^{\circ}$ from the linear receptors, the differential gain and phase between the receptors were adjusted to accomplish polarization calibration. 
We determined the rotation measure, and obtained a new (RM). 
The new RM value  is $77.6 \pm 1.9 \, rad m^{-2}$, which is consistent with the result given by \citet{2019MNRAS.484.3646S}.
With the newly measured RM, the polarized pulse profiles of this pulsar are corrected.  
After the above preprocessing, for the 40-minute tracking observation of PSR J2007$+$0910, a total of 5094 single pulses are obtained.
Figure \ref{fig_1} shows a portion of the pulse stack of this pulsar.
Each pulse profile has 1024 phase bins.
Finally, the single pulse was analyzed by using the PSRSALSA package\citep{2016A&A...590A.109W}, which is available online for free download \footnote{https://github.com/weltevrede/psrsalsa}.
Using the PSRSALSA to analyze the fluctuation spectrum of single pulses, the drifting parameters of subpulses of this pulsar are measured.

\section{The behaviors of single pulse and subpulse}\label{sec_3}

\subsection{The energy distribution of single pulse}\label{sec_3.1}

\begin{figure} 
 \centering
\includegraphics[width=0.48\textwidth]{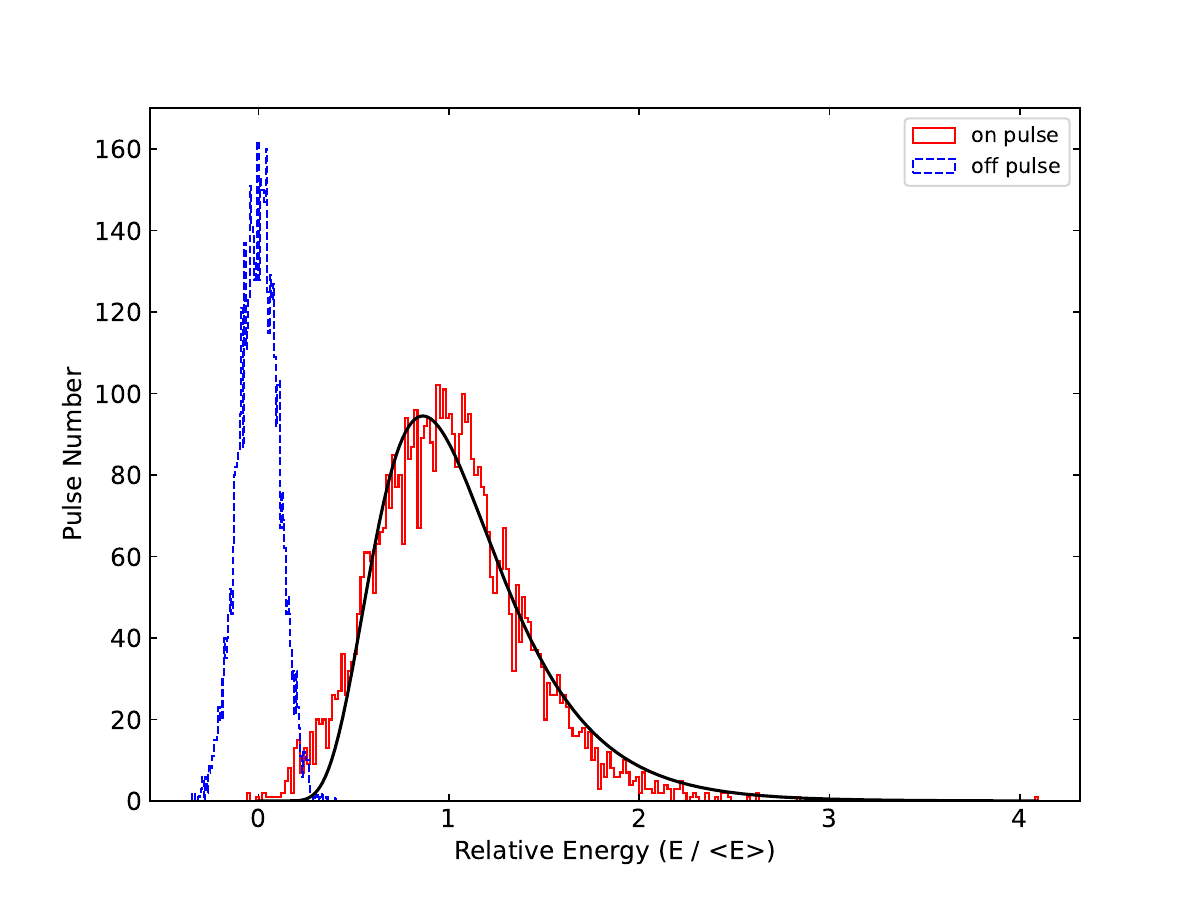}
 \caption{Pulse energy histograms for the on-pulse (red) and off-pulse (blue) windows. The energies are normalized by the mean on-pulse energy. The black curve line is a log-normal distribution that has been fitted to the on-pulse energy histogram.
 \label{fig_2}} 
\end{figure}

We measured the integrated flux density of each single pulse of PSR J2007+0910 to obtain its energy distribution. The on-pulse energy of each pulse can be obtained by summing the intensities of the pulse phase bins within the on-pulse window of the mean pulse profile. Similarly, a region of the same width as the on-pulse window is used to estimate the off-pulse energy. The energy distributions for on-pulse and off-pulse are shown in Figure\,\ref{fig_2}, where the energies have been normalised by the mean on-pulse energy. 
As can be seen, the shape of the off-pulse energy histogram is Gaussian and is centered on zero.
The on-pulse energy histogram shows a very clear peak at the mean pulse energy. The lack of a bimodal shape in the energy distribution of the on-pulse indicates that there is no obvious nulling for this pulsar.

We fitted the energy distribution of the on-pulse region using a log-normal function:
\begin{equation}
  Y(E) = \frac{A}{E \sigma \sqrt{2 \pi}} exp\left[-\frac{(ln(E) - \mu)^2}{2 \sigma^2}\right]
  \label{equ_1}
\end{equation}
where $A$, $\mu$ and $\sigma$ are normalization factor, logarithmic mean and the standard deviation of the distribution, respectively. 
We employed a Kolmogorov-Smirnov (KS) hypothesis test for the energy distribution of on-pulse, which is used to test whether or not it can be described as a log-normal distribution. The obtained $p$ value for KS hypothesis test is 0.42 (greater than the threshold value of 0.05), which implies that the energy distribution of on-pulse is consistent with a log-normal distribution. The black curve in Figure\,\ref{fig_2} is the result of fitting Equation\,\ref{equ_1}.

\subsection{Subpulse drifting}\label{sec_3.2}

PSR J2007$+$0910 is pulsar which exhibits a significant subpulse drift behavior. Its drift behaviour has been reported by several researchers \citep[e.g. ][]{2005MNRAS.363..929C, 2023MNRAS.520.4562S}. Its single pulse, however, has not been analysed in detail. 
Here, we used FAST to make a high-sensitivity observation, and found that there is more than one drift mode in the subpulse of PSR J2007$+$0910.
For example, the right panel of Figure\,\ref{fig_1} shows that the subpulse drift rate suddenly decreases at the $\sim 1850th$ pulse.
But not only that, but interestingly, its subpulses sometimes exhibit stationary states. For example, the left panel of Figure\,\ref{fig_1} shows that the subpulse exhibits almost no drift behaviour at pulse numbers of $\sim 250$ and $\sim 450$.
The frequent switching of various drift modes and the coexistence of drift and non-drift behavior make the emission phenomena of this pulsar even more fascinating.
In this section, we presented a detailed analysis of the drifting subpulses phenomenon of PSR J2007$+$0910.

\subsubsection{Drift Characteristics}\label{sec_3.2.1}

The Longitude Resolved Fluctuation Spectra \citep[LRFS, ][]{1970Natur.227..692B} is an effective method for identify any periodically repeating pattern as a function of pulse longitude. But it cannot be used to identify whether the pulsed emission drifts in pulse longitude from pulse to pulse. And this shortcoming can be effectively solved using the method of the the two-dimensional fluctuation spectrum \citep[2DFS, ][]{2002A&A...393..733E}.
2DFS can simultaneously obtain the pattern repetition frequency along the pulse longitude axis (i.e., $P_3$) and the separation between subpulses of a successive drift bands in the pulse longitude (i.e., $P_2$)

\begin{figure} 
 \centering
 \includegraphics[width=0.45\textwidth]{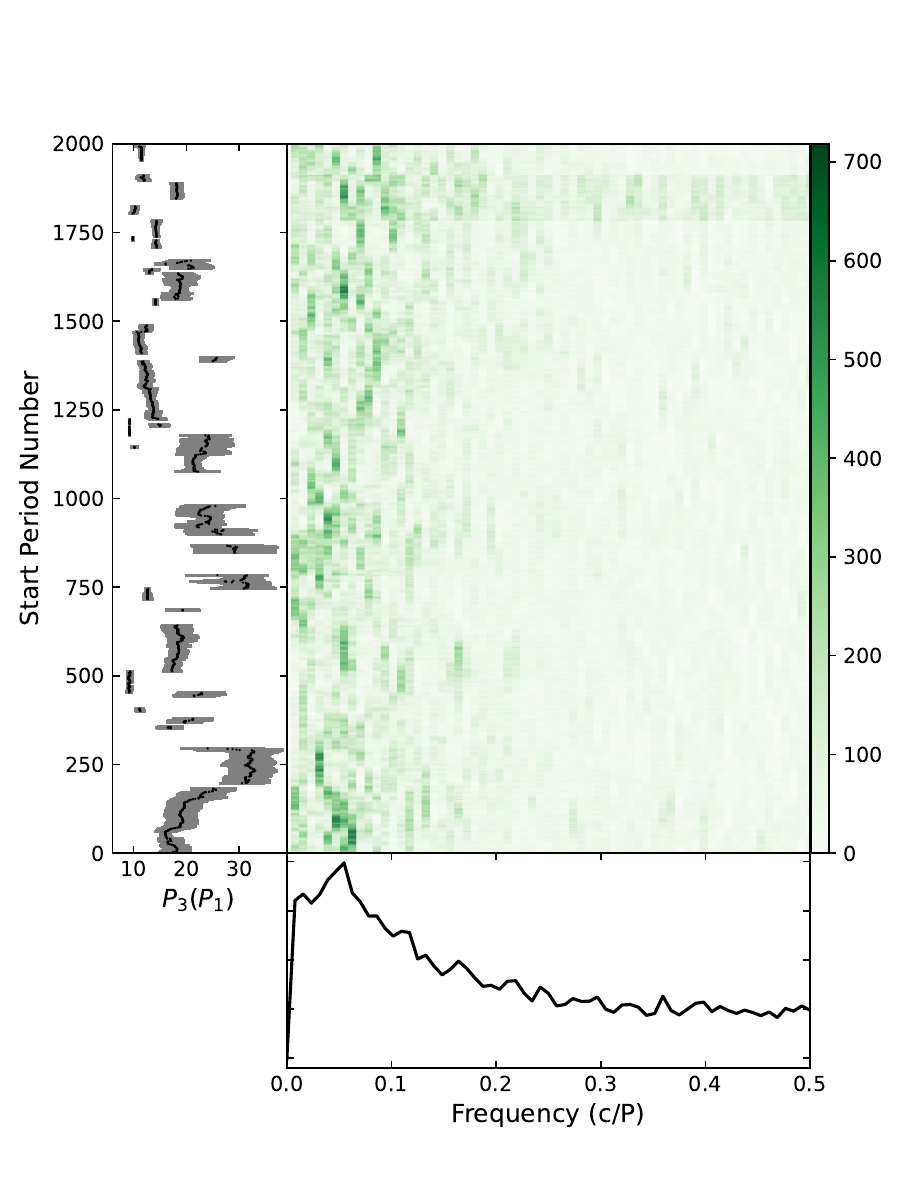}
 \caption{Part of the Sliding two-dimensional fluctuation spectrum (S2DFS) for this observation of J2007$+$0910. The spectral power of 128 consecutive single pulses were determined. The starting point was then shifted by one period and the process was repeated until the end of this observation. The left panel shows the $P_3$ values measured at each shift of the starting point, and the bottom panel shows the time-averaged fluctuation spectrum. The uncertainties correspond to $1 \sigma$.
 \label{fig_3}} 
\end{figure}

2DFS was employed to study the drift characteristics in PSR J2007$+$0910. We used the PSRSALSA package \citep{2016A&A...590A.109W} to obtain the results of 2DFS for the specified pulse sequence.
However, the switching between different drift modes makes the results of the spectral analysis of the entire observation not to show a pure drift feature. Also the process of averaging leads to a loss of some information, making the detection and characterization of any time-related changes very difficult.
The Sliding 2DFS (S2DFS) proposed by \citet{2009A&A...506..865S} was used to solve this problem as a way to find the time interval in which the drifting behavior is stable. 
We calculated the 2DFS of intervals of 128 single pulses shifting the starting point by one period \citep{2009A&A...506..865S}. $P_3$ and $P_2$ values are determined by fitting a Gaussian near the major peak in the 2DFS. 
We show the variation of the 2DFS as a function of the start period, the results are shown in Figure\,\ref{fig_3}. 
The left panel of Figure\,\ref{fig_3} obviously shows the instability of the $P_3$ values with time.
The $P_3$ value fluctuates from a few pulse periods to more than thirty pulse periods. For example, between the starting pulses $\sim 0-300$, the $P_3$ value increases from $\sim 20~P$ to $\sim 30~P$. And between the starting pulses $\sim 1250 - 1500$, the $P_3$ value changes again to $\sim 17~P$. Drift mode switching is the main cause of drastic variations in $P_3$ values. 
The bottom panel shows the time-averaged fluctuation spectrum.

\begin{figure} 
 \centering
 \includegraphics[width=0.45\textwidth]{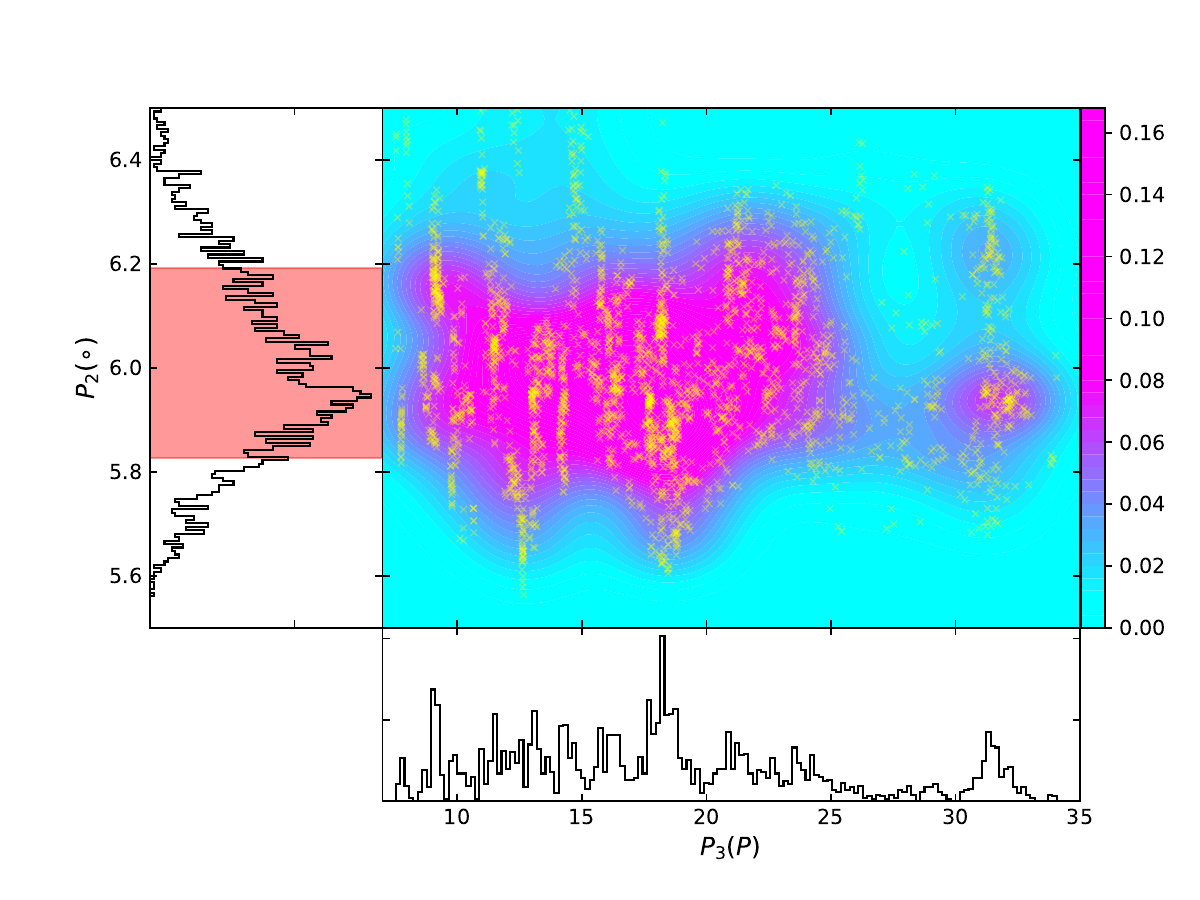}
 \caption{Density distribution of $P_2$ vs. $P_3$ for PSR J2007$+$0910. The density distribution of the data is represented in contours of different colours. The left panel and the bottom panel give the distribution of $P_2$ and $P_3$, respectively. The $P_3$ distribution indicates the presence of multiple drift modes, and the $P_2$ distribution indicates that the $P_2$ values for these drift modes are essentially the same.
 \label{fig_4}} 
\end{figure}

Based on the results of S2DFS, we obtained the density distributions of $P_2$ vs. $P_3$ for PSR J2007$+$0910, and the results are shown in Figure\,\ref{fig_4}. 
In contrast to pulsars with only one drift mode, the P3 values of PSR J2007$+$0910 do not show a normal distribution, but are scattered between $7\,P$ and $35\,P$, which is due to the drift mode switching of this pulsar.
As shown in the bottom panel of Figure\,\ref{fig_4}, $P_3$ exhibits multiple peaks, in which $P_3 = 8.5, 18, 32\,P$ are very distinct. This means that there are multiple drift modes in this pulsar. The frequent switching of different modes causes the dispersion of its $P_3$ value.
However, since the duration of some drift modes is short sometimes, this makes it possible that the modulation period obtained using S2DFS is provided by two or even three drift modes. 
That is, although $P_3$ exhibits multiple peaks, these peaks are not to be considered to be distinct drift modes, but only as intuitive appearances of the presence of multiple drift modes. 
2DFS analysis of pulse sequences with long burst duration can be used to determine accurate $P_3$ values for different drift modes, which will be described in detail in section\,\ref{sec_3.2.2}. 
It is worth noting that the distribution of $P_2$ measurements is always the same for all cases. $P_2$ is independent of the type of drift mode. This has also been found in other observational studies of multiple drift mode pulsars \citep[e.g. ][]{2022MNRAS.509.2507R, 2022ApJ...934...23S, 2023MNRAS.520.1332Z}. 
Via averaging the $P_2$ values, we get $P_2 = (6.01 \pm 0.18) ^\circ$, where the uncertainty corresponds to the standard error. As shown in the red area of the left panel of Figure\,\ref{fig_4}.


\subsubsection{Boundaries and characteristics of the modes}\label{sec_3.2.2}

The discovery of multiple drift modes in PSR J2007$+$0910 has added another member to the population of pulsars with multiple drift types. The high sensitivity observations provide us with more details, which  is prime importance for understanding the emission from this multi-drift pulsar. 
It's important to mention that mode switching is always immediate and some bursts of modes don't last very long. 
This presents a challenge in identifying the different modes.

In this paper, we employed the phase-averaged power spectrum (PAPS) proposed by \citet{2005A&A...440..683S} to determine the boundaries of the different modes.
We first selected a pulse sequence with obvious subpulse drifting. Then we obtained the square of the absolute value of the Fourier transform of the flux for each fixed pulse phase over the entire pulse window of the sequence and obtained the PAPS by averaging the resulting transforms over the pulse phase. Finally, the beginning and ending of the pulse sequence are adjusted to obtain the highest signal-to-noise ratio of the peaks, where the peak value of PAPS divided by the root mean square of the remaining values is considered as the signal-to-noise ratio of PAPS. Meanwhile the final result is obtained after careful checking. 
This method has been successfully used to separate the drift modes of some multiple drift pulsars \citep[e.g. ][]{2022MNRAS.509.2507R, 2022MNRAS.509.4573J, 2023MNRAS.520.1332Z}.
Based on the identification results of PAPS, we obtained the 2DFS of each pulse sequence using the PSRSALSA package \citep{2016A&A...590A.109W}. Analyzing their drift characteristics in this way.
However, 2DFS analysis can be limited by the length of the pulse sequence. For shorter pulse sequences, it is rather challenging to obtain a reliable result.

\begin{figure} 
 \centering
 \includegraphics[width=0.48\textwidth]{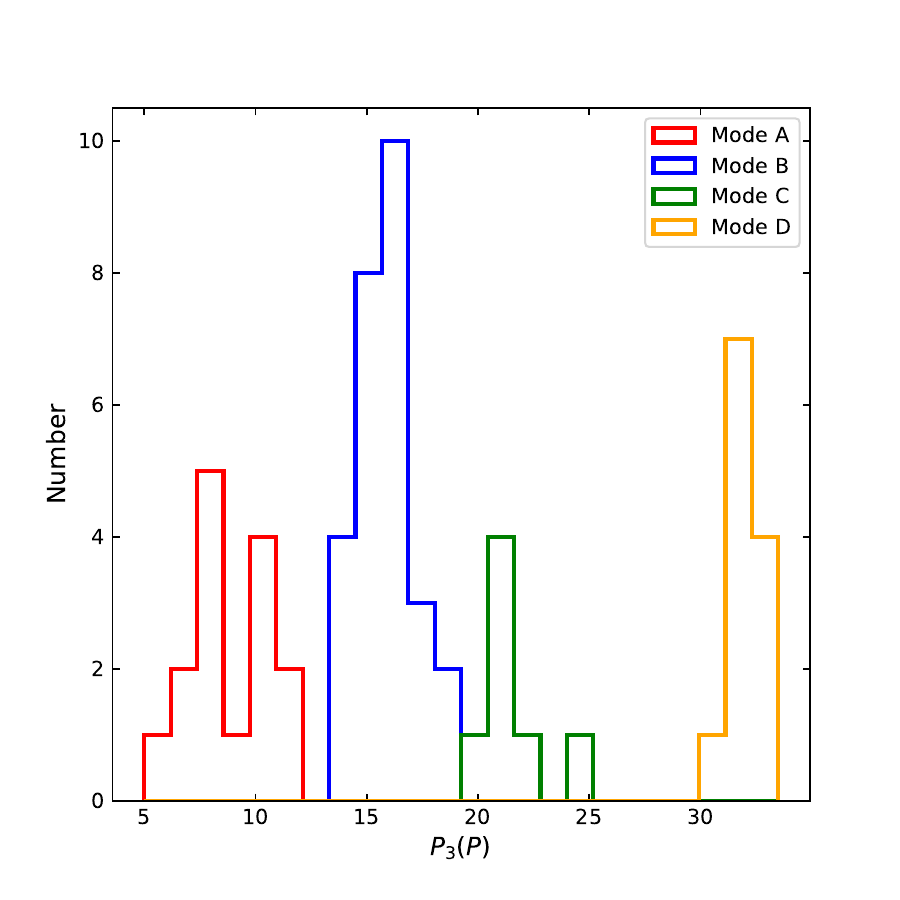}
 \caption{Distribution of $P_3$ values for drift modes with longer burst durations. Different modes are displayed in different colours.
 \label{fig_5}} 
\end{figure}

\begin{figure*}
    \centering
    \label{fig:2}
    \includegraphics[width=4.2cm,height=5.3cm]{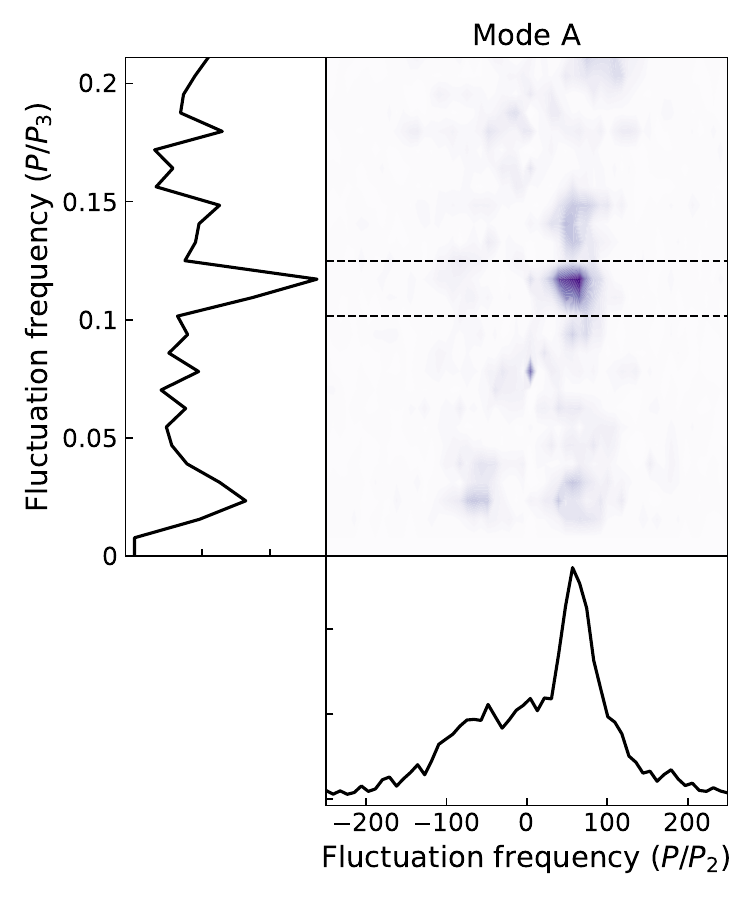}
    \includegraphics[width=4.2cm,height=5.3cm]{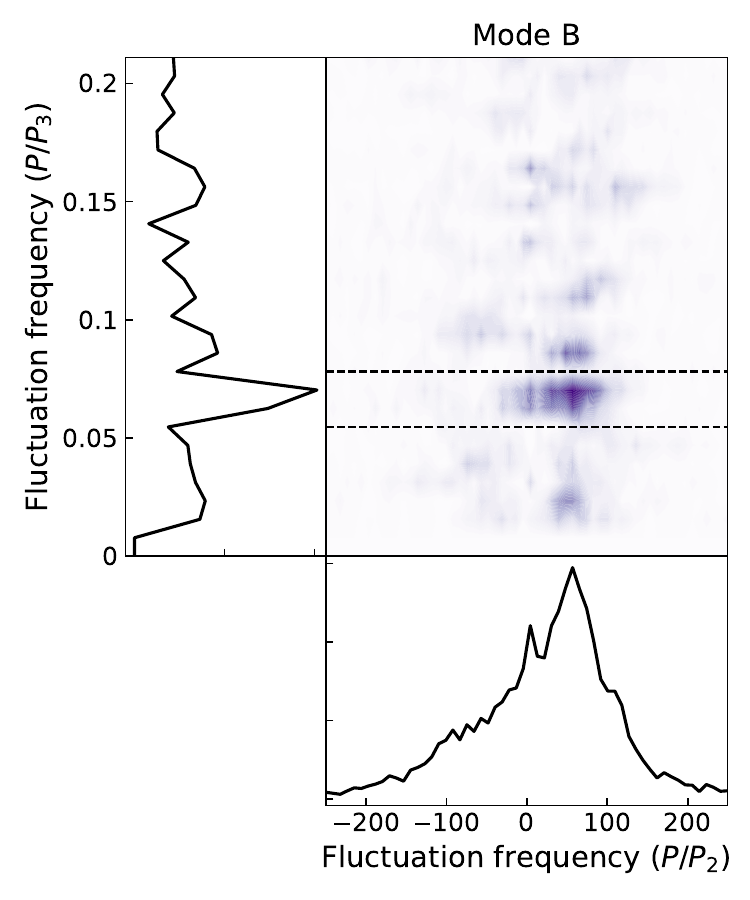}
    \includegraphics[width=4.2cm,height=5.3cm]{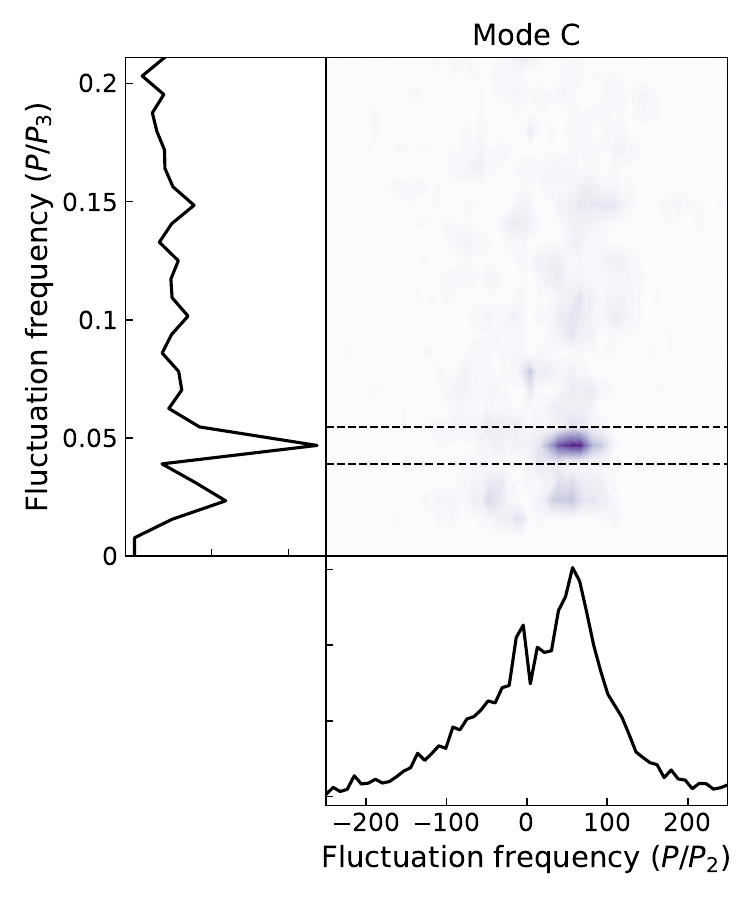}    
    \includegraphics[width=4.2cm,height=5.3cm]{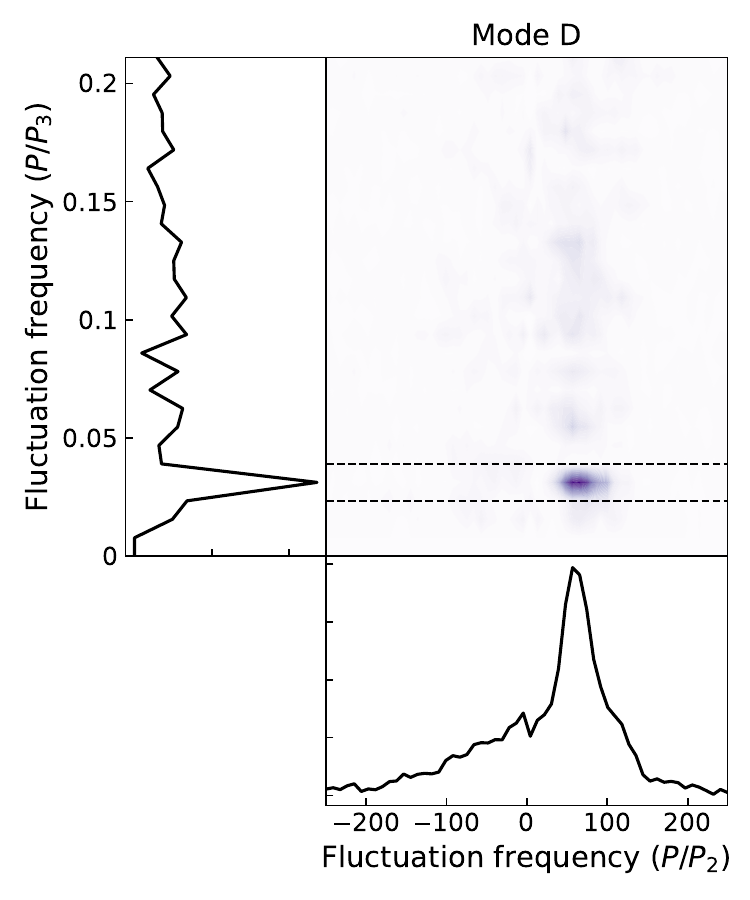}  
    \caption{Examples of 2DFS for drift modes A, B, C and D. The horizontal integrated power and vertical integrated power in the 2DFS were shown in the left and bottom panels, respectively. For the left panel, the power at frequencies below 0.01 c/p is set to zero. The darkest value corresponds to the strongest integrated power. It is easy to see the similarities and differences in the structure of the features in the 2DFS of each modes.
    \label{fig_6}}
\end{figure*}

\begin{figure*} 
 \centering
 \includegraphics[width=0.95\textwidth]{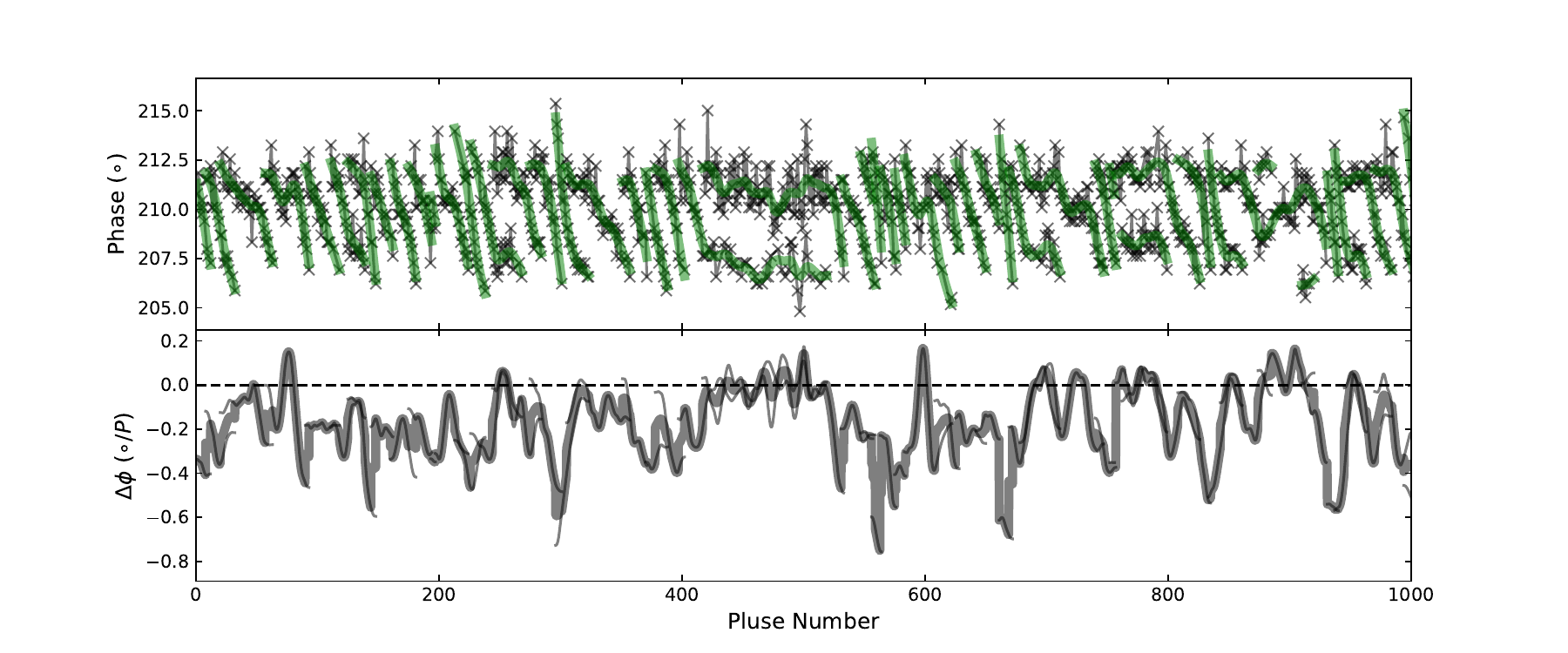}
 \caption{Drift patterns and drift rates for single pulses of J2007$+$0910. In the top panel, the black crosses indicate the locations of the subpulses and the green line corresponds to fits to the drift bands using the cubic smoothing spline method. In the panels below, the black solid line shows the drift rate for an individual track, while the grey solid line shows the average drift rate. Only 1000 consecutive single pulse sequences are shown here. 
 \label{fig_7}} 
\end{figure*}

Therefore, we used this method only for longer duration pulse sequences with subpulse drift behaviour, for obtaining their Drift characteristics.
Figure\,\ref{fig_5} show the distribution of $P_3$ for these pulse sequence. The result show that PSR J2007$+$0910 exhibits at least four different drift modes.
They are named as modes A, B, C, and D, respectively, according to the distribution of $P_3$. 
Figures\,\ref{fig_6} is an examples of 2DFS analyses for these drift modes. 
In Figures\,\ref{fig_6}, the horizontal integrated power and vertical integrated power in the 2DFS are shown in the left and bottom panels, respectively. 
For the left panel, the power at frequencies below $0.01 c/p$ is set to zero.  
We can clearly see that their vertically integrated powers are essentially the same. However, the horizontal integrated power is different for the different modes, with the peaks are $P_{3, A} = 8.5 P$, $P_{3, B} = 14.2$, $P_{3, C} = 21.3$ and $P_{3, D} = 32.0$, respectively.
In addition to this, we observe two emission modes with non-drift behavior, classified as modes $E_1$ (double peaks) and $E_2$ (single peaks) according to the morphology of the pulse profile.
In general, J2007$+$0910 exhibits at least six modes in our observations. 

\begin{figure} 
 \centering
 \includegraphics[width=0.48\textwidth]{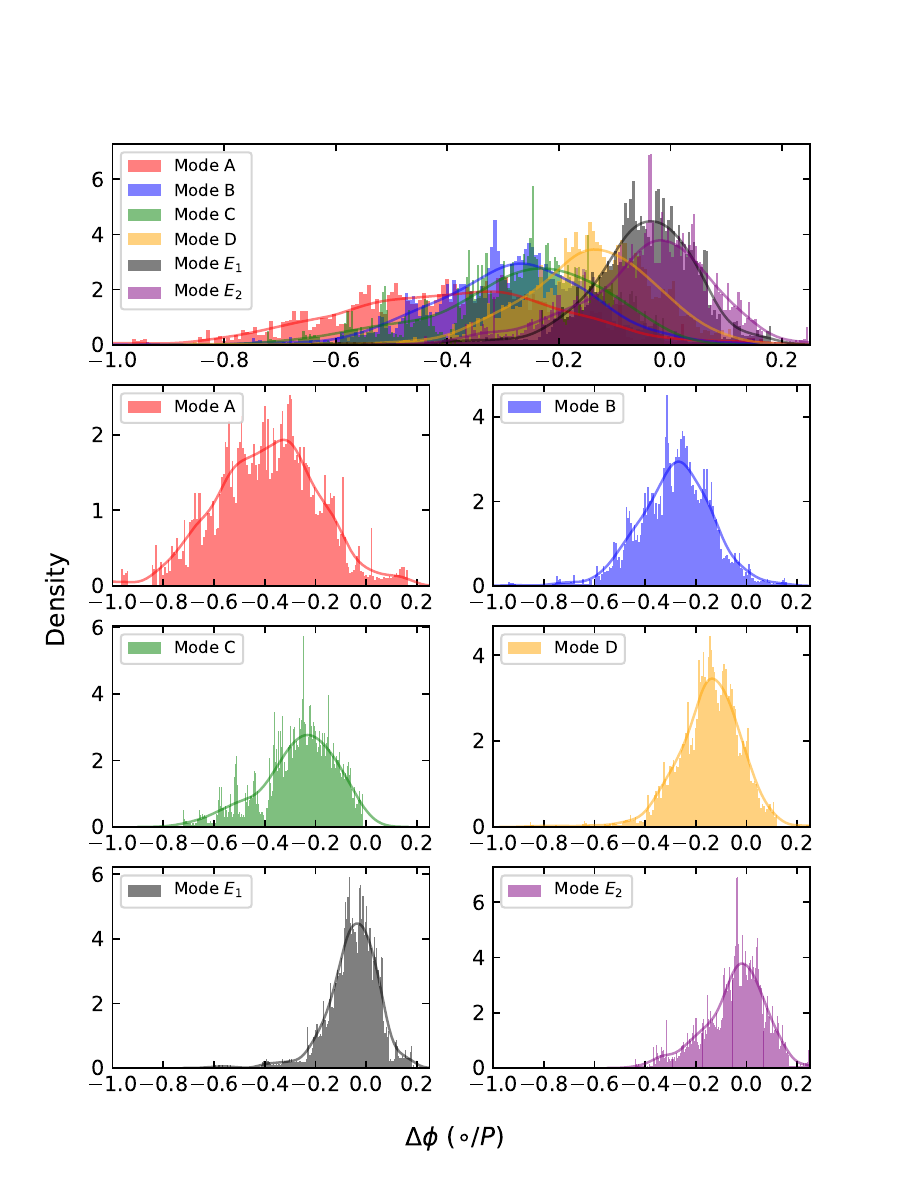}
 \caption{Distribution of drift rate $\Delta \phi$ for different modes. The most top panel of these shows the combined distribution for all modes. The lower six panels show the distribution of drift rate $\Delta \phi$ for individual modes.
 \label{fig_8}} 
\end{figure}

For shorter duration pulse sequences, we further distinguish which drift mode they belong to by studying the difference in the drift rate with each pulse in the drift band. To calculate the drift rate, we first obtained the subpulse locations by convolving the data from a single pulse with a Gaussian function having the mean width of the subpulses. The peaks in the convolution are used to determine the longitude of the subpulse. Then, the trajectory of each drift band is estimated using the triple smoothing spline method. Finally, the drift rate is obtained by calculating the gradient of the fitted spline function at each pulse number for each drift band.
Figure\,\ref{fig_7} shows an example of a portion of the observation session. Its top panel shows the drift trajectory of one thousand consecutive pulses, where the black crosses indicate the locations of the subpulses and the green line corresponds to fits to the drift bands using the cubic smoothing spline method. In the bottom panel, the black solid line shows the drift rate for an individual track. 
As some pulses contain two subpulses, yielding two measurements of the drift rate for that pulse number, we estimated their drift rates by averaging the contribution of each drift band. The solid grey line in the bottom panel of Figure\,\ref{fig_7} indicates the average drift. 
The change in the drift rate at mode switches is evident from Figure\,\ref{fig_7}. for example, at a pulse number of $\sim 300$ the drift mode switches to mode A.  
And the emission mode at $\sim 420$ is changed from mode B to mode $E_1$.
A cumulative and modal histogram of individual drift rates obtained by taking a gradient of the smoothing spline fit of driftbands at each pulse number can be seen in Figure\,\ref{fig_8}. 
The top panel of Figure\,\ref{fig_8} shows the distribution of all drift rates, where each colour corresponds to the different modes classification of subpulse of PSR J2007$+$0910, and the lower panels show the distribution of drift rates for its all six modes.
The results show that modes $E_1$ and $E_2$ have almost zero peak drift rate, which is an indication that they do not exhibit drift characteristics. For the four modes with drift behaviour, they have different drift rate peaks, with mode D having the largest and mode A the smallest. 

\subsubsection{Drift Modes}\label{sec_3.2.3}

\begin{figure} 
 \centering
 \includegraphics[width=0.48\textwidth]{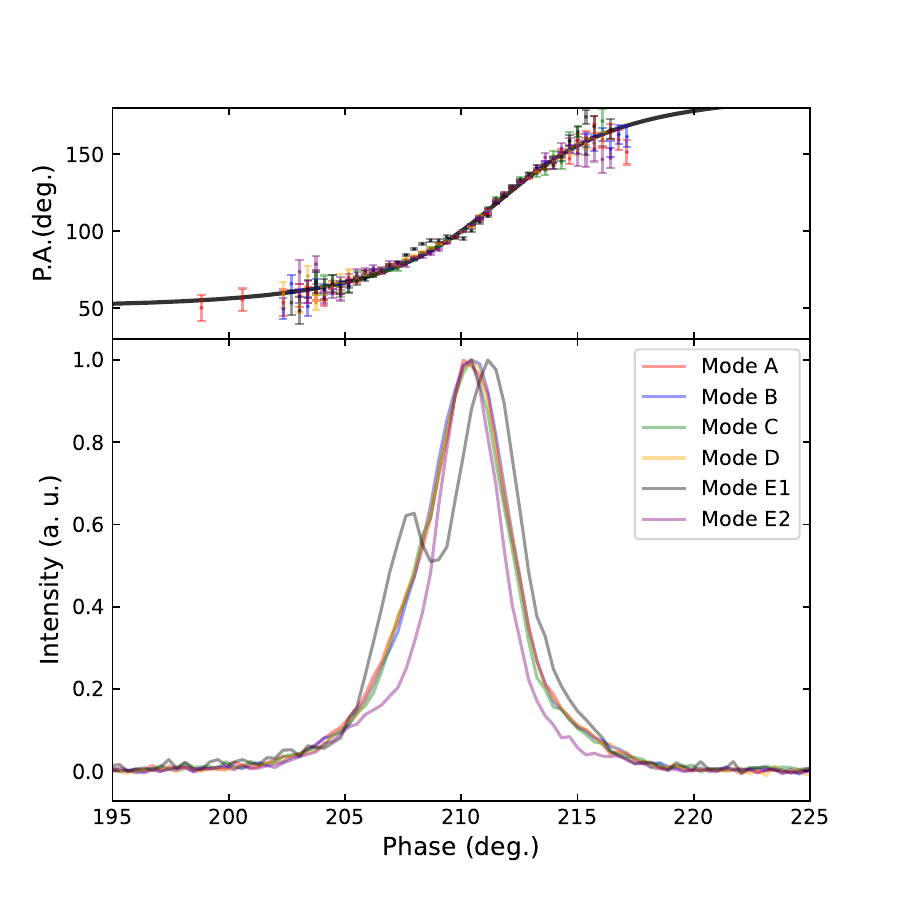}
 \caption{Polarisation profiles of different drift modes. The lower panel shows the integrated pulse profile, where the pulse profile has been normalised. The top panel shows the P.A. as a function of pulse longitude for all modes.
 \label{fig_9}} 
\end{figure}

\begin{table*}
\centering
\caption{The results of each parameters are listed for each modes. \label{table_1}}
{\begin{tabular}{ccccccccccc}
\hline\hline
Mode & $P_3$ & $P_2$ & Number of & Number of & $T_{min}$ & $T_{max}$ & $W_{50}$ & $W_{10}$  & $<L> / I$ & $<V> / I$ \\
 & ($P$) & ($\circ$) & sequences & pulses & ($P$) & ($P$) & ($\circ$) & ($\circ$)  & ($\%$) & ($\%$) \\
\hline
 A & 8.7 $\pm$ 1.6 & 5.8 $\pm$ 0.5 & 31 & 953 & 7 & 99 & 4.3 $\pm$ 0.4 & 10.8 $\pm$ 0.4 & 38.2 $\pm$ 2.3 & 6.3 $\pm$ 1.6 \\
 B & 15.8 $\pm$ 1.2 & 6.2 $\pm$ 0.5 & 34 & 1423 & 12 & 94 & 4.3 $\pm$ 0.4  & 10.5 $\pm$ 0.4 & 38.5 $\pm$ 1.8 & 6.1 $\pm$ 1.1 \\
 C & 21.6 $\pm$ 1.3 & 6.1 $\pm$ 0.5 & 8 & 480 & 36 & 75 & 4.3 $\pm$ 0.4 & 10.2 $\pm$ 0.4 & 37.9 $\pm$ 3.5 & 4.8 $\pm$ 2.0 \\
 D & 32.3 $\pm$ 0.9 & 6.1 $\pm$ 0.7 & 25 & 1109 & 15 & 146 & 4.4 $\pm$ 0.4 & 10.6 $\pm$ 0.4 & 38.1 $\pm$ 2.1 & 5.6 $\pm$ 1.5 \\
 $E_1$ & $-$ & 6.0 $\pm$ 0.4 & 26 & 506 & 6 & 55 & 5.9 $\pm$ 0.4 & 10.8 $\pm$ 0.4 & 35.5 $\pm$ 3.2 & 8.6 $\pm$ 2.0 \\
 $E_1$ & $-$ & 5.9 $\pm$ 0.6 & 21 & 623 & 5 & 60 & 3.2 $\pm$ 0.4 & 9.1 $\pm$ 0.4 & 38.9 $\pm$ 3.0 & 5.4 $\pm$ 1.6 \\  
\hline\hline	
\end{tabular}}
\end{table*}

We separated all the modes, and found thatPSR J2007+0910 shows six different modes in our observations, four of which exhibit clear drift behavior, while the other two modes show a quiescent subpulse structure.
For all six modes a brief description is given below:

\textbf{Mode A: }
This mode has significant drift behavior (As shown in the pulse sequence from $\sim$ 1810th to 1840th single pulse in the right panel of Figure\,\ref{fig_1}). The $P_3$ distribution for this mode is defined from between $5\,P$ to $13\,P$. 
The calculated average value is $P_3 = 8.4~P$. 
This mode has been reported previously by \citet{2005MNRAS.363..929C} in observations at a frequency of 430~MHz, and their results show $P_3 = 11.8 P$, it is consistent with $P_3$ of mode A given by this paper. 
Our observations show that its occurrence is 18\% and that most bursts are of relatively short duration, with the shortest lasting only a few periods, and the longest bursts approaching a hundred periods.
This paper believes that this mode represents a variety of indistinguishable drift modes. 
The shorter durations and nearly identical $P_3$ values present a difficulty to separate them in more detail (see Section\,\ref{sec_5} for details).

\textbf{Mode B: }
Relative to the previous mode, mode B exhibits a slower drift rate as well as longer period modulation. 
As the sequence from $\sim$1865th to $1900$th single pulses in the right panel of Figure\,\ref{fig_1} showing, the average $P_3$ and $P_2$ values are $15.8~P$ and $6.2^\circ$, respectively. 
This mode has not been reported in previous studies. Mode B has the highest occurrence in our observations, which exists for 28\% of the entire observation time.

\textbf{Mode C: }
The mean $P_3$ of this model is larger than that of modes A and B (As the $\sim$ 400th pulse shown in the left panel of Figure\,\ref{fig_1}). The average $P_3$ is $21.6~P$ . The same results as those observed by \citet{2023MNRAS.520.4562S} using MeerKAT. However, the uncertainty in their results is large, probably because they did not separate the modes, and the $P_3$ value is the result of the integrated power superposition of all the modes. 
This mode occurs less frequently in our observations, but none of its durations are too short. It has a minimum burst duration of $36$ pulse periods, which is the longest of all the modes.

\textbf{Mode D: }
This mode has the longest modulation period of all the drift modes (As the $\sim$310th to $\sim$370th pulses showing in the left panel of Figure\,\ref{fig_1}).
The average $P_3$ value of this mode is $32.3~P$. 
This mode has not been reported by previous studies. In our observation, about 22\% of single pulses are observed in this mode.

\textbf{Mode $E_1$: }
Mode $E_1$ does not exhibit subpulse drift behaviour, and the integrated pulse profile shows two peaks, which is due to the line-of-sight sweeping over the two sparks in the radiation window. 
The subpulse location does not move in phase with time, and it looks as if the sparks above the polar cap have stopped rotating. 
However, we prefer to think that this is due to the fact that after one period the spark point moved exactly the distance between the spacing of the two spark points. 
This will be discussed in more detail in Section\,\ref{sec_5}. 
The duration of this mode is mostly short, with a minimum of only a few pulse periods. It is worth noting that it has a larger $W_{50}$ (the full width of the pulse profile at 50\% of the pulsar amplitude) than that of other modes. 
While $W_{10}$ (the full width of the pulse profile at 10\% of the peak) 
is basically the same as that in other modes (see figure\,\ref{fig_9}).

\textbf{Mode $E_2$: }
Like the  mode $E_1$,  this mode has not drifting subpulse behavior. 
The difference between the modes $E_1$  and $E_2$ is that the single pulse profile of  mode $E_2$ shows a single peak morphology. 
The mechanism of generating this mode is similar to that of mode  $E_1$.
The difference in generation mechanisms between the modes $E_1$ and $E_2$ is that the number of spark points in the emission window of mode $E_2$ is only one.
This mode only exhibited for 10\% of our observation time. 
The line of sight sweeps over only one spark, resulting in the mode having the smallest pulse width in all the modes.

We summarise a summary of the various parameters for the different modes in Table\,\ref{table_1}. 
The second and third columns are $P_3$ and $P_2$, respectively.
Note that all the values of $P_3$ and $P_2$ are averages of the corresponding modes, and the standard deviations are taken as uncertainties. 
The number of pulse sequences, the number of single pulses and the maximum and minimum values of burst duration for each mode are shown in the fourth column to seventh column. 
The eighth and ninth columns are the width of pulse at 50\% and 10\% of peak of each mode, respectively.
It is worth noting that the integrated pulse widths of the modes with drift behavior are highly consistent, while the pulse widths of the modes without drift behavior are wider or narrower relative to that of drifting modes.
The last two columns show the fractions of linear polarisation and the fractions of circular polarization, respectively. 
The fractions of linear and circular polarisation of the six mode are almost Self-consistent within the error range.

The integrated pulse profiles and PA information for the six modes are shown in Figure\,\ref{fig_9}. 
It is found that the integrated profiles of the four drifting modes are basically the same. 
In contrast, the integrated profile of non-drifting mode $E_1$ is wider, and show a double peak morphology.
While the morphology of the integrated profile of $E_2$ is the same as that of the four drifting modes, showing a clear single peak morphology, but its profile is narrower than that of the four drifting modes.
It is interesting that the PA swings shown in the upper panel of Figure\,\ref{fig_9} are almost the same, which means that these modes, including drifting and non-drifting modes, are generated from the same region of the pulsar magnetosphere.


\section{The magnetosphere configuration of PSR J2007$+$0910}\label{sec_4}

\begin{figure*}
 \centering
 \includegraphics[width=0.45\textwidth]{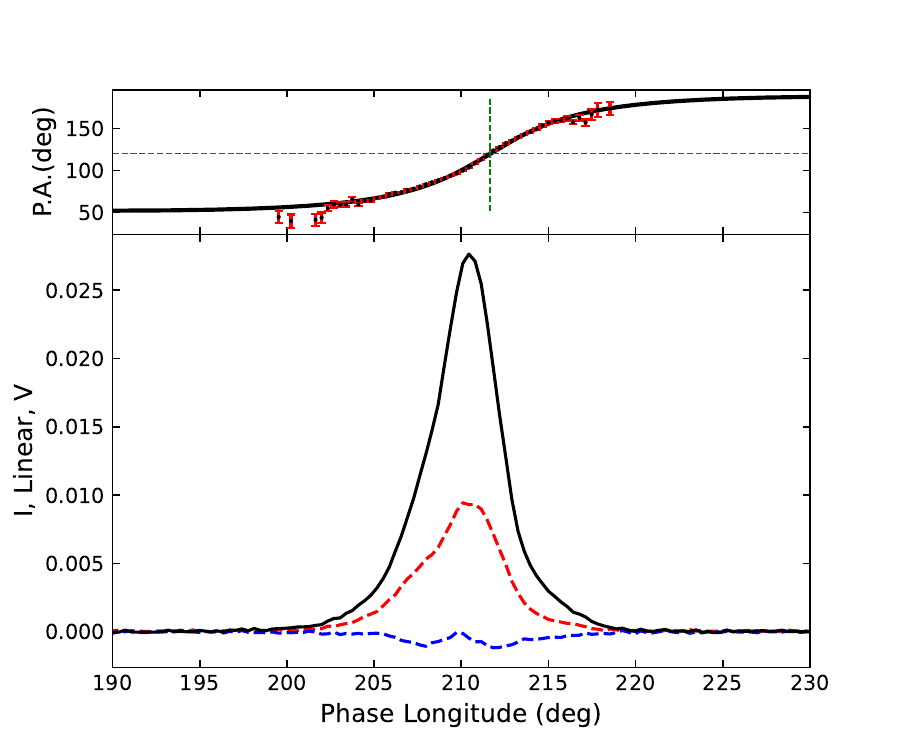}
 \includegraphics[width=0.45\textwidth]{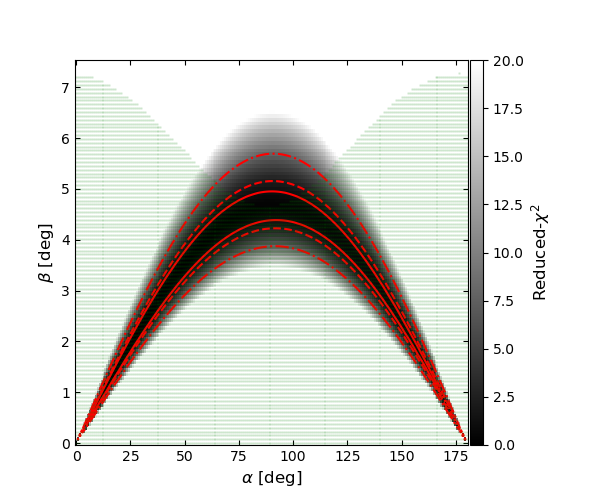}
 \caption{Left: the polarization profile at 1250~MHz. Where the total power, total line polarization and circular polarization are represented by black solid line, red dashed line, and blue dashed line, respectively. The black dots, in the top panel, represent the linear polarization angles along with the best-fit curve from the RVM fit shown as the black curve. Right:  the results of fitting an RVM curve for each ($\alpha$, $\beta$) combination. The reduced chi-squared ($\chi^2$) of the fit is shown as the gray scale, with the darkest value corresponding to the best fit. The red contour lines represent $1 \sigma$, $2 \sigma$, and $3 \sigma$ confidence boundaries. They are represented by solid, dashed and dotted dashed lines respectively. The green shaded regions are the “allowed” viewing geometries, which are constraints arising from the estimated emission height and observed profile width. 
 \label{fig_10}}
\end{figure*} 

We obtained the integrated pulse profile and polarization information of J2007$+$0910 at center frequency of 1250~MHz by accumulating the single pulses shown in the left panel of Figure\,\ref{fig_10}.
In the meantime, to remove the influence of the pulsar ionosphere and the interstellar medium during propagation, we had corrected the integrated pulse profile using the RM values determined by the RMFIT package.
The results are shown in the upper left panel of Figure\,\ref{fig_10}, where the total intensity, total linear polarization and circular polarization are represented by the black solid line, red dashed line and blue dashed line, respectively.
The pulse profile exhibits moderate linearly polarized with a fractional line polarization of $(37.7 \pm 0.9)\%$.
PSR J2007$+$0910 exhibits the same edge depolarization as the conal-single pulsar \citep{1983ApJ...274..333R}. 
The upper left panel of Figure\,\ref{fig_10} shows the results of the position angle (PA) of its linear polarization. In order to obtain the information of the viewing geometry,  we fitted the observed PA curve with the Rotating Vector Model \citep[RVM, ][]{1969ApL.....3..225R}: 
\begin{equation}
    \tan(\psi-\psi_0)=\frac{\sin \alpha \sin(\phi-\phi_0)}{\sin (\alpha + \beta) \cos \alpha - \cos (\alpha + \beta) \sin \alpha \cos(\phi-\phi_0)}
\end{equation}
where $\alpha$ and $\beta$ are the magnetic inclination angle (the angle between the spin and magnetic axes) and the impact angle (the angle between the line of sight and magnetic axes), respectively. $\psi$ is the PA at a pulse longitude $\phi$, and $\psi_0$ and $\phi_0$ are the phase offsets for polarization position angle and pulse rotational phase, respectively.
The relativistic motion of the emission region relative to the observer as the region corotates with the pulsar causes aberration and retardation \citep[A/R; ][]{2004ApJ...607..939D, 2015MNRAS.446.3356R}.
This effect causes a delay in the pulse phase of the inflection point of the PA curve predicted by RVM relative to the the location of the fiducial plane inferred from the intensity profile.
\begin{equation}
  \Delta \phi = \phi_0 - \phi_{fid} = \frac{8 \pi h_{em}}{P c}
\end{equation}
where $P$ is the spin period of the pulsar, $c$ is the speed of light, $h_{em}$ is the emission height \citep{1991ApJ...370..643B}. $\phi_0$ can be estimated by RVM fitting. $\phi_{fid}$ can be obtained based on the pulse profile morphology. 
The upper limit of our observed delayed is $1.63^\circ$, which means that the emission height is calculated below $156 \pm 4~km$. 
According to the method described by \citet{2015MNRAS.446.3356R}, the half-opening angle, $\rho$, of the beam of radio emission can be estimated by the value of the emission height by the following equation:
\begin{equation}
  \rho = A + arctan\left(\frac{tanA}{2}\right)
\end{equation}
where $A = arcsin(\sqrt{2 \pi h_{em} / (P c)} )$ is the  angular radius of the open-field-line region. 
The half-opening angle of the conical emission beam centered on the magnetic axis is related to $W_0$ ( the rotational phase range for which the line of sight samples the open-field-line-region) \citep{1984A&A...132..312G}:
\begin{equation}
  cos \rho = cos\alpha cos(\alpha + \beta) + sin\alpha sin(\alpha + \beta) cos\left(\frac{W_0}{2}\right)
\end{equation}
As the fiducial plane does not necessarily correspond to the center of the observed pulse, which means that $W_0$ is not necessarily the same as the observed pulse width. We assumed that $W_0$ is equal to twice the difference in phase between the fiducial plane position and the farthest pulse edge from it, where the pulse edge is the point at which the emission is $10\%$ of the peak intensity. 
Thus according to equations (4) - (5) one can constrain $\alpha$ and $\beta$, and the results are shown in the green area of the right panel of Figure\,\ref{fig_10}.
The combination of the RVM fit and the A/R effect to constrain $\alpha$ and $\beta$ allows us to obtain a specific combination of $\alpha$ and $\beta$ (as shown in Figure\,\ref{fig_10}). The results show that the magnetic inclination, $\alpha$, is practically unconstrained and we can only conclude that $\beta \le 7^\circ$.

%


\section{discussions}\label{sec_5}

In this paper, we reported  multi-drifting behavior of subpulses of PSR J2007$+$0910 with the FAST observation at a centre frequency of 1250~MHz. 
Our results shows that the subpulse of this pulsar exhibits multi-drifting mode, and no null pulses are detected in the single pulse.
The multi-drifting behavior makes the pulsar  one of the promising candidates for studying the generation mechanism of multi-drifting subpulse behavior of pulsar. Although previous studies have reported subpulse drifting behaviour of PSR J2007$+$0910, the sensitivity limitations of telescope have not revealed the existence of a multi-drifting subpulse. 
The high-sensitive observations of FAST can  provide more details of single pulse emission of pulsars, which is crucial for our discovery of the multi-drifting subpulse behavior of this pulsar.
For PSR J2007$+$0910, we found six subpulse modes in the single pulse emission of this pulsar for the first time, in which four of them show drifting behavior, two of them arr non-drifting modes. 
We named the four drifting modes as modes A, B, C and D according to the distributions of $P_3$ values. 
Although two of these modes (modes A and C) have been reported in previous studies \citep{2005MNRAS.363..929C, 2023MNRAS.520.4562S}, as above described, no detailed mode separation was carried out in previous works, this makes the estimated integrated intensity is the summation of the integrated intensity of all modes so that the calculated modulation period will differ from reality.  
We carried out a mode separation, and obtained more accurate modulation periods for each drifting mode (As shown in Table\,\ref{table_1}). 
In this section, the corresponding discussions and conclusions will be given.

In the traditional carousel model, it is assumed that the sparks above the polar cap are equally spaced in magnetic azimuth, and the subpulse drift phenomenon is observed as the line of sight cuts through the slightly rotating sparks. 
The theoretical value of $P_4$ is predicted by the RS model\citep{1975ApJ...196...51R}, as follows,
\begin{equation}
 \overline{P}_{4} = 5.7 \times \left( \frac{P_1}{s} \right)^{-3/2} \left( \frac{\dot{P}}{10^{-15}} \right)^{1/2}
\label{equ_3}
\end{equation}
The results is $\overline{P}_{4} \approx 10.6$. 
Actually, the emission mode of pulsar can be affected by magnetospheric interactions \citep{2022MNRAS.514.4046W}. Sampling rate of once a period makes it possible for observed drifts to be aliased \citep{2003A&A...399..223V}. The presence of the aliasing effect implies that the apparent drift rate $\Delta \phi$ is a confection of the larger true drift $\Delta \phi^{'}$ and sampling rate, i.e. that the spark drift in two consecutive samples is comparable to the $P_2$ value. 
From the equation, $P_3 = P_3^{'} / (P_3^{'} - 1)$ , where $P_3^{'} = P_2 / \Delta \phi ^{'}$ is the true repetition rate. It can be seen that just a small change in $\Delta \phi^{'}$ or $P_2$ can cause a significant change in the $P_3$ value of the apparent drift \citep{2013MNRAS.433..445R}. 
If aliasing exists, the observed $P_3$ obeys this relation, $P_3 = P_4 / (|n - k\,P_4|)$, where the unit of $P_3$ and $P_4$ is the pulse period, $k = [n / P_4]$ is the aliasing order, in which the square brackets denote rounding to the nearest integer. 
For pulsars with multi-drifting modes, the different $P_3$ for each mode can be caused by the variations of the number of sparks points $n$ or $P_4$, or both. 
But the rotational speed of the carousel is determined by the magnetic and electric fields near the pulsar's surface, which cannot be easily changed in magnitude and or direction (i.e. $P_4$ is not easily changed).
Therefore different drift behaviors may be occurred by simply changing $n$ and keeping $P_4$ constant \citep{2013MNRAS.433..445R}. 
If it is assumed that $k$ is the same for all drifting modes and the number of spark points in neighbouring modes is an equidistant sequence, then the $P_3$ of each mode will also be harmonically related \citep{1981A&A...101..356W}.
Based on this, we considered the first-order aliasing effect (i.e., $k = \pm 1$) and studied all the modes of PSR J2007$+$0910. 
A representative first-order solutions are shown in Table\,\ref{table_2}.

\begin{table}
\centering
\caption{The carousel model expectation of each modes for different $n$, assuming that $P_3$ is a first-order blending of $P^{'}_{3}$. \label{table_2}}
{\begin{tabular}{cccccccccc}
\hline
\multicolumn{4}{c}{$k = 1$} && \multicolumn{4}{c}{$k = -1$}\\
 \cline{1-4}\cline{6-9}&\\
n & $P_3$ & $P^{'}_{3}$ & $P_4$ && n & $P_3$ & $P^{'}_{3}$ & $P_4$ & Mode\\
 & ($P$) & ($P$) & ($P$) && & ($P$) & ($P$) & ($P$) & \\
\hline
 64 & ($\infty$) & 1.00 & && 65 & ($\infty$) & 1.00 &  & $E_1, E_2$\\
 63 & (64.0) & 1.02 & && 66 & (65.0) & 0.98 &  & -\\
 62 & 32.0 & 1.03 & && 67 & 32.5 & 0.97 &  & D\\
 61 & 21.3 & 1.05 & && 68 & 21.7 & 0.96 &  & C\\

 60 & 16.0 & 1.07 & && 69 & 16.3 & 0.94 &  & B\\
 $59^*$ & 12.8 & 1.08 &  && $70^*$ & 13.0 & 0.93 &  & \\
 $58^*$ & 10.7 & 1.10 & 64.0 && $71^*$ & 10.8 & 0.92 & 65.0  & \\
 $57^*$ & 9.1 & 1.12 &  && $72^*$ & 9.3 & 0.9 &  & \\
 $56^*$ & 8.0 & 1.14 &  && $73^*$ & 8.1 & 0.89 &  & A\\
 $55^*$ & 7.1 & 1.16 &  && $74^*$ & 7.2 & 0.88 &  & \\
 $54^*$ & 6.4 & 1.19 &  && $75^*$ & 6.5 & 0.87 &  & \\
 $53^*$ & 5.8 & 1.21 & && $76^*$ & 5.9 & 0.86 &  & \\
 $52^*$ & 5.3 & 1.23 &  && $77^*$ & 5.4 & 0.84 &  & \\
\hline	
\end{tabular}}
\end{table}

As shown in the results of Table\,\ref{table_2}, we successfully predicted all observed modes. 
As expected, the fitted P3 values have a distinctly harmonic properties. Taking multiples of $64.0 (k=1)$ or $65.0 (k=-1)$, the observed modes are $1/2$ (Mode D), $1/3$ (Mode C), $1/4$ (Mode B), and $1/8$ (the mid-range for Mode A). 
This is not for the first time that this periodicity has been found in pulsars with multiple drift modes.
As can be seen from the results in Table\,\ref{table_2}, a theoretical mode ($n = 63/66$) with a slower drift rate  should exist for this pulsar, which has a drift rate of $P_3 \pm 64/65$. However, it is worth noting that we did not observe this theoretical mode in our observations.
We suggested that this may be of short duration and indistinguishable from $E_1$ and $E_2$. 
It can be seen that a small change in the true modulation period can leads to a significant change in the observed $P_3$. 
In the last column of Table\,\ref{table_2} we gave the corresponding drift modes. 
The results show that when $k=1$, the number of sparks of modes B, C and D are $n_B = 60$, $n_C = 61$ and $n_D = 62$, respectively.
Noting that if the generating circle of sparks has a circulation time of $P_3^{''}$ pulse periods, the intrinsic repetition rate to be $P_3^{'} = P_3^{''} / n$ \citep{2013MNRAS.433..445R}.  
This implies that 64 subbeams will produce infinite $P_3$, i.e. the true drift $\Delta \phi ^{'}$ is almost equal to $P_2$. In this particular case, the pulsar will exhibit sustained non-drifting emission with a narrow $P_2$. 
At the same time, since the separation distance of the subbeam is smaller than the pulse emission window, we can observe radio emission from two or one subbeam in the emission window. 
These two possibilities correspond to observed modes $E_1$ and $E_2$ of this paper, respectively.
Depending on the predictions, 52 to 59 subbeams will produce the observed mode A (as describe in Section 3 of this paper, we categorised the drift modes with period modulation between $5\,P$ and $13\,P$ as mode A). 
Since the $P_3$ values produced from these sparks are very close to each other and our observations show that they are mostly short in duration (e.g., Table\,\ref{table_1}), distinguishing between all the modes included in mode A is extremely challenging. 
Similarly when $k = -1$, the corresponding $n$, $P_3$ and $P_3^{'}$ of all modes are shown in Table\,\ref{table_2}. 
Where 70 to 77 subbeams will produce mode A, while modes B, C and D have spark numbers of $n_B = 69$, $n_C = 68$ and $n_D = 67$, respectively. 
The non-drifting modes $E_1$ and $E_2$ are generated when the spark number are 65.
The predicted number of sparks increases with the aliasing orders.
Whereas the approximate solution of $P_4$ represents the fastest possible carousel rotation speeds consistent with the $P_3$, the transformation of the aliasing order does not lead to a significant change in it.
At present, some studies suggest that the multiple drift modes may be caused by the first-order or seceond-order alias of a faster drift of subbeams equally spaced around the cones \citep{2013MNRAS.433..445R, 2019ApJ...883...28M, 2022MNRAS.509.2507R}.  
If higher aliasing orders are assumed, smaller $P_3^{'}$ variations can produce the observed $P_3$.

Several studies on pulsars with multiple drift modes have found that the drift modes occur in sequences of increasing drift rates \citep[e.g.][]{1970ApJ...162..727H, 2005MNRAS.357..859R}. 
In order to determine the sequence of drifting modes in the mode changes of PSR J2007$+$0910, we statistically analyze the sequence of mode changes in this observation.
The statistics show that mode A or mode B usually occurs after mode $E_1/E_2$, and then the drift mode switches to mode $E_1/E_2$ again. Its drift mode switching exhibits a characteristic modal sequence pattern: drift mode from $E_1/E_2$ to $A$ or $B$ and then switching back to $E_1/E_2$. 
In addition to this, we note that there are also frequent switches between modes A or B and mode D, but mode D usually occurs before modes A or B. After that, the drift mode will be switched to mode $E_1/E_2$ instead of mode D. That is, drift mode switching exhibits another characteristic modal sequence pattern: drift mode from $D$ to $A$ or $B$ and then switching back to $E_1/E_2$. 
More specifically, the sequence of drifting modes in the mode changes of this pulsar seems to conform to a general rule which states that the mode sequences always observed from slower to faster drifts and then restart with the mode at the slower drift rate.
Certainly, the opposite also occurs, though, such as mode A to mode B switching, or mode $E_1/E_2$ to mode D switching. This means that the drift modes can also also occur in a sequence that decreases the drift rate, but the probability of this happening is relatively small in our observations.
It is worth noting that modes C and D do not seem to switch to each other, at least not in our observations. This needs to be verified with longer observations.


\section{SUMMARY}\label{sec_6}

We carried out a detailed study of the drifting subpulse behavior in PSR J2007$+$0910 at central frequency 1250\,\rm{MHz} with the FAST. 
Based on the ultra-high sensitivity of the FAST, we discovered some interesting subpulse drift phenomena, i.e., the pulse emission of PSR  J2007$+$0910 shows multi-drifting subpulse modes, and no nulls are found in this pulsar. 
Via a K-S hypothesis test, we found that the energy distribution of the on-pulse region follows a lognormal distribution. For the pulse emission, there are six different emission modes are found in this pulsar. Among them, Four of these modes show obvious subpulse drifting behavior (i.e., modes A, B, C and D), and two modes (modes $E_1$ and $E_2$) show stationary subpulse structures. 
These was not reported in previous studies. 
For the four subpulse drifting modes, their drift characteristics are analysed  by using the PSRSALSA package. 
The results shows that $P_3$ of modes A, B, C and D are $P_{3,A} = 8.7 \pm 1.6$, $P_{3,B} = 15.8 \pm 1.2$, $P_{3,C} = 21.6 \pm 1.3$ and $P_{3,D} = 32.3 \pm 0.9$, respectively. 
It is noteworthy that the $P_2$ of these modes are almost identical, the value of $P_2$ is $\sim 6.01 \pm 0.18^\circ$.
We considered first-order aliasing effects and found that all six modes of this pulsar may be generated from aliasing effects. The different drift rates and $P_3$ values can be obtained by changing only the number of sparks in the carousel, the corresponding solutions are shown in Table\,\ref{table_2}.
The results suggest that the multi-drifting and non-drifting modes of this pulsar may be caused by an aliasing effect of under-sampling rate.
At the same time, we counted the sequences in the mode changes. We found that the mode switching is not random, and mode sequences are almost always observed from slower to faster drifts.
The number of pulsars with multi-drifting modes may be increased with the successive high-sensitivity observation of the FAST, this will provide an opportunity to systematically study the multi-drifting phenomenon, which is useful for understanding radiation mechanism of pulsar.

\section*{Acknowledgements}
This work is supported by the National SKA Program of China (Nos. 2022SKA0130100, 2022SKA0130104),the National Natural Science Foundation of China (No. 12273008), the Natural Science and Technology Foundation of Guizhou Province (No. [2023]024), the Foundation of Guizhou Provincial Education Department (No. KY (2020) 003), 
the Academic New Seedling Fund Project of Guizhou Normal University (No. [2022]B18)  and the Scientific Research Project of the Guizhou Provincial Education (Nos. KY[2022]137, KY[2022]132, ZK[2022]304), the Major Science and Technology Program of Xinjiang Uygur Autonomous Region (No.2022A03013-4)
This work uses the data from the FAST telescope (Five-hundred-meter Aperture Spherical radio Telescope). FAST is a Chinese national mega-science facility, built and operated by the National Astronomical Observatories, Chinese Academy of Sciences.

\section*{Data Availability}
The data underlying this article will be shared on reasonable request to the corresponding author.



\bibliographystyle{mnras}
\bibliography{MN} 



\bsp	
\label{lastpage}
\end{document}